\documentclass[a4paper,fleqn]{cas-dc}

\usepackage[numbers]{natbib}

\def\tsc#1{\csdef{#1}{\textsc{\lowercase{#1}}\xspace}}
\tsc{WGM}
\tsc{QE}
\tsc{EP}
\tsc{PMS}
\tsc{BEC}
\tsc{DE}

\hypersetup{
	pdftitle={D2A U-Net: Automatic Segmentation of COVID-19 Lesions from CT Slices with Dilated Convolution and Dual Attention Mechanism},
	pdfsubject={Engineering},
	pdfauthor={Xiangyu Zhao, Guanglei Zhang},
	pdfkeywords={COVID-19, Segmentation, Deep Learning, Attention Mechanism, Dilated Convolution},
}

\begin{document}
\let\WriteBookmarks\relax
\def\floatpagepagefraction{1}
\def\textpagefraction{.001}

\author[1]{Xiangyu Zhao}
\author[1]{Peng Zhang}
\author[1]{Fan Song}
\author[1]{Guangda Fan}
\author[1]{Yangyang Sun}
\author[1]{Yujia Wang}
\author[1]{Zheyuan Tian}
\author[1]{Luqi Zhang}
\address[1]{School of Biological Science and Medical Engineering, Beihang University, Beijing, China}

\author[1,2]{Guanglei Zhang}[orcid=0000-0002-2617-9673]
\cormark[1]
\ead{guangleizhang@buaa.edu.cn}
\address[2]{Beijing Advanced Innovation Center for Biomedical Engineering, Beihang University, Beijing, China}

\cortext[cor1]{Corresponding author}

\title[mode = title]{D2A U-Net: Automatic Segmentation of COVID-19 Lesions from CT Slices with Dilated Convolution and Dual Attention Mechanism}

\begin{abstract}
\noindent \emph{Background and Objective}: Coronavirus Disease 2019 (COVID-19) has caused great casualties and becomes almost the most urgent public health events worldwide. Computed tomography (CT) is a significant screening tool for COVID-19 infection, and automated segmentation of lung infection in COVID-19 CT images will greatly assist diagnosis and health care of patients. However, accurate and automatic segmentation of COVID-19 lung infections remains to be challenging. In this paper we propose a \textit{ dilated dual attention U-Net} (\textit{D2A U-Net}) for COVID-19 lesion segmentation in CT slices based on dilated convolution and a novel dual attention mechanism to address the issues above. 

\noindent \emph{Methods}: We introduce a dilated convolution module in model decoder to achieve large receptive field, which refines decoding process and contributes to segmentation accuracy. Also, we present a dual attention mechanism composed of two attention modules which are inserted to skip connection and model decoder respectively. The dual attention mechanism is utilized to refine feature maps and reduce semantic gap between different levels of the model. 

\noindent \emph{Results}: The proposed method has been evaluated on open-source dataset and outperforms cutting-edges methods in semantic segmentation. Our proposed \textit{D2A U-Net} with pretrained encoder achieves a Dice score of 0.7298 and recall score of 0.7071. Besides, we also build a simplified \textit{D2A U-Net} without pretrained encoder to provide a fair comparison with other models trained from scratch, which still outperforms popular U-Net family models with a Dice score of 0.7047 and recall score of 0.6626. 

\noindent \emph{Conclusion}: Our experiment results have shown that by introducing dilated convolution and dual attention mechanism, the number of false positives is significantly reduced, which improves sensitivity to COVID-19 lesions and subsequently brings significant increase to Dice score. Significance: Our proposed method improves segmentation performance on COVID-19 lesions in CT slices, and can be regarded as a potential AI-based approach utilized in diagnosis and prognosis of COVID-19 patients.
\end{abstract}


\maketitle

\section{Introduction}

COVID-19 pandemic caused by SARS-nCov-2 continues to spread all over the world \cite{wang2020novel}, and most of the countries have been affected in this unprecedented public health event. By August 2020, more than 23.75 million of cases of COVID‑19 have been reported and more than 810,000 died \cite{coronavirus.jhu.edu} of COVID-19 infection. Due to the strong infectivity of SARS-nCov-2, identification of people infected by COVID-19 is significant to cut off the transmission and slow down virus spread. Reverse transcriptase-polymerase chain reaction (RT-PCR) is considered as the gold standard of diagnosis \cite{wang2020detection} for its high specificity, but it is time-consuming and laborious. Also, the capacity of RT-PCR tests can be rather insufficient in less-developed regions, especially during the pandemic. Computed tomography (CT) imaging is one of the most commonly used screening methods to detect lung infection and has proved to be efficient in the diagnosis and follow-up prognosis of COVID-19.

\begin{figure}[h]
	\centering
	\includegraphics[width=0.44\textwidth]{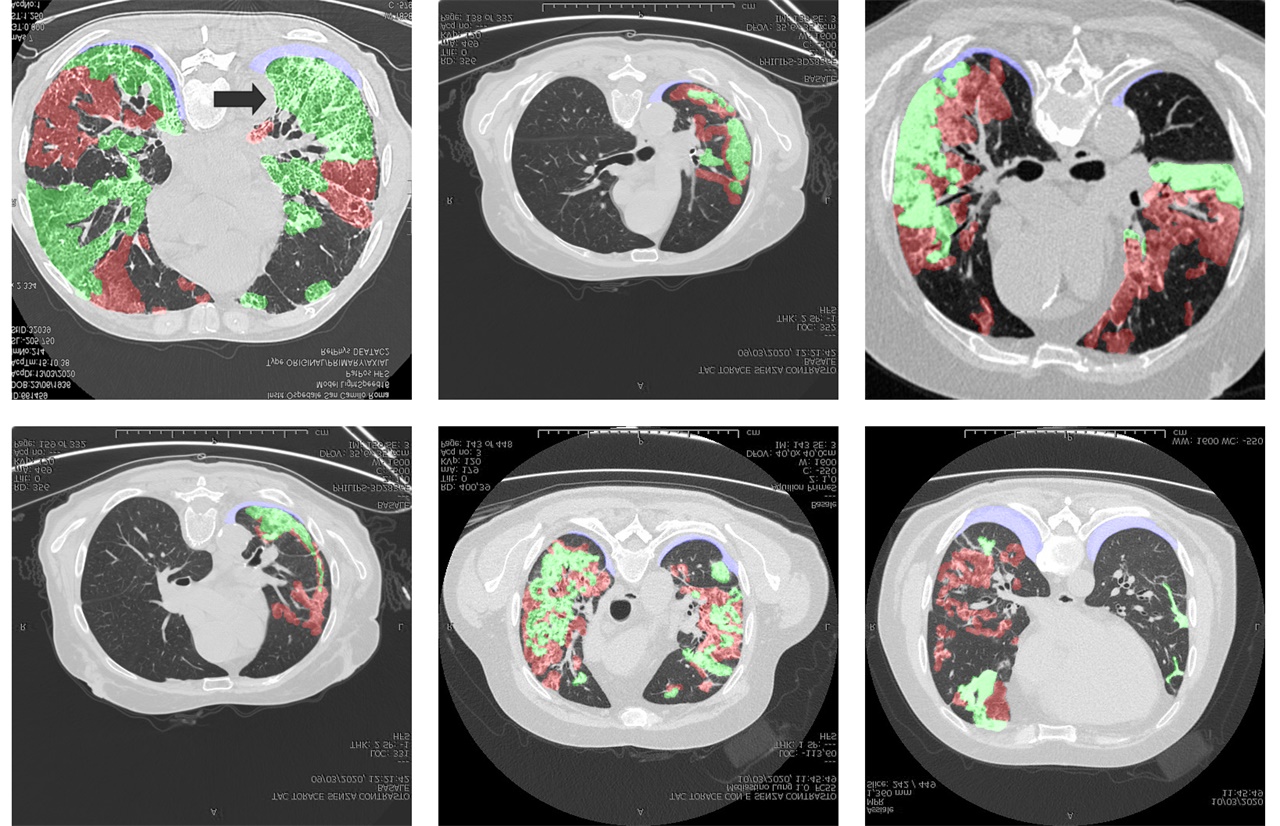}
	\caption{Example of COVID-19 CT slices, where the red, green and blue masks denote the ground glass, consolidation and pleural effusion respectively. The images are collected from \cite{covid19ctseg}.} 
	\label{fig:covidct}
\end{figure}

Compared with chest X-ray images, CT imaging is more sensitive, especially in the early stage of infection. Ground glass pattern is the most common finding in COVID-19 infections, usually in the early stage, while pulmonary consolidation can be observed in the later stage. Pleural effusion can also be observed in pathological CT slices. These typical features of COVID-19 lung infection are shown in Fig. \ref{fig:covidct}.

Thus, chest CT imaging is regarded as a convenient, fast and accurate approach to diagnose COVID-19. The evaluation of localization and geometric features of infection area could provide adequate information of disease progress and help doctors make better treatment \cite{lei2020ct} \cite{ng2020imaging} \cite{pan2020time}. However, manual annotation of infection regions is a time-consuming and laborious work. Also, the annotation made by radiologists can be subjective and biased due to individual experience and personal judgements.

Recently, a number of deep learning systems using convolutional neural networks (CNNs) have been proposed to detect COVID-19 infection. For instance, Wang and Wong \cite{wang2020covid} have developed a COVID-Net to perform ternary classification between healthy people, COVID-19 patients and people infected with other pneumonia in chest X-ray images, which achieves an overall accuracy of 93.3 $\%$. In terms of deep learning systems for CT imaging, Zhou and Canu \cite{zhou2020automatic} have proposed an automatic network facilitated with attention mechanism to segment infection area in CT slices. Fan et. al \cite{fan2020inf} developed an Inf-Net and corresponding semi-supervision algorithm to perform CT segmentation. Zheng et al. \cite{zheng2020deep} proposed a weakly-supervised deep learning method to detect COVID-19 in CT volumes. Xi et al. \cite{ouyang2020dual} presented a dual-sampling attention network to diagnose COVID-19 from community acquired pneumonia. However, detecting lung infection caused by COVID-19 in CT images remains to be challenging. As infection regions vary in shape, position and texture, and the boundaries with normal tissues can be rather blurred, which add to the difficulty in COVID-19 detection and limit model performance, especially in terms of recall score.

To address the issues above, we proposed a \textit{dilated dual attention U-Net} (\textit{D2A U-Net}) framework to automatic segment lung infection in COVID-19 CT slices. Since infected tissues can be hardly distinguishable with normal tissues, we introduce a dual attention mechanism consisting of a \textit{gate attention module} (GAM) and a \textit{decoder attention module} (DAM) to refine feature maps and produce more informative feature representation. The proposed GAM is utilized by fusing features and semantic-rich gate signals to refine skip connections. The proposed DAM is introduced to the decoder of the network to improve model decoding quality and better segment blurred infected tissues. As COVID-19 infection varies in position and size, we utilize dilated convolution with different dilation rate in the model decoder to obtain larger receptive fields and balance the segmentation on both large and tiny objects, which thus provides better segmentation results. Such refinement improves segmentation recall score and thus provides better segmentation results.

The paper is organized as follows: Section \ref{sec:related} offers a review of related works on CT segmentation. Section \ref{sec:methods} describes the overview of this work and details our model. Section \ref{sec:experiments} presents the details of our experiments and provides both quantitative and qualitative segmentation results. Section \ref{sec:conclusion} discusses the proposed method and concludes our work.

\section{Related Works}
\label{sec:related}

\begin{figure*}[h]
	\centering 
	\includegraphics[width=0.95\textwidth]{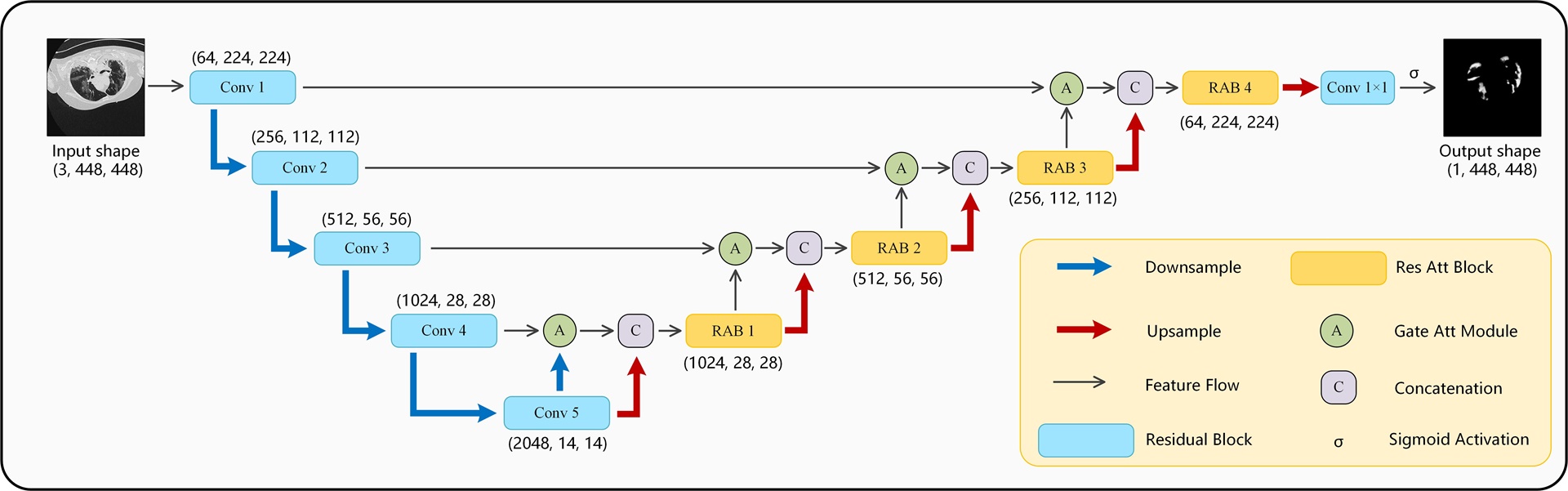}
	\caption{The proposed \textit{D2A U-Net} architecture with a ResNeXt-50 (32$\times$4d) backbone, which takes a CT slice as input and outputs infection region predictions. See \ref{sec:netarchitecture} for details.} 
	\label{fig:netimage} 
\end{figure*}

In this section, we will go through 4 types of most related works, which includes chest CT segmentation, attention mechanism, dilated convolution and AI-based COVID-19 segmentation systems.

\subsection{Chest CT Segmentation}
Chest CT imaging is one of the most popular screening methods for lung disease diagnosis\cite{kamble2020review}. Segmentation of organs and lesions provides crucial information for disease diagnosis and prognosis. However, manual segmentation remains time-consuming and laborious and subjective error is inevitable, thus automatic CT segmentation gains much popularity in the research fields. Recent researches upon automatic segmentation mainly focus on utilizing machine learning techniques. Related works most feature a pixel-wise classifier to infer from extracted features to make predictions. For example, Mansoor et. al \cite{mansoor2014generic} proposed a texture-based feature classifier for pathological lung segmentation in CT images. Yao et. al \cite{yao2011computer} utilized texture analysis and support vector machine to segment infections in lung tissues. These algorithms have realized automatic segmentation in chest CT images but several issues remain unsolved, including subjective bias in feature extraction and difficulties in segmenting nodule regions. Deep learning algorithms feature powerful fitting capacity and require no laborious preprocessing. Most cutting-edge segmentation algorithms are based on deep learning approaches. For example, Shaziya et. al \cite{shaziya2018automatic} used U-Net to segment lung tissues in chest CT scans. Zhao et. al \cite{zhao2018lung} proposed a fully convolutional neural network with multi-instance and conditional adversary loss for pathological lung segmentation.

\subsection{Attention Mechanism}
Attention plays an important role in human perception and visual cognition \cite{corbetta2002control}. One significant property in human perception is that humans hardly process visual information as a whole. Instead, humans usually process visual information recurrently, where top information is utilized to guide bottom-up feedforward process \cite{mnih2014recurrent}. Inspired by this principle, attention mechanism has been widely used in computer vision, especially in image classification \cite{hu2018squeeze} \cite{woo2018cbam} \cite{wang2017residual}. Related algorithms typically refine feature maps in spatial dimension, channel dimension or both. For example, Hu et al. \cite{hu2018squeeze} introduced a Squeeze-and-Excitation module, where global average pooling is performed on input features to produce channel-wise attention. Woo et. al \cite{woo2018cbam} proposed a convolutional block attention module (CBAM) to introduce a fused attention consisting of channel attention and spatial attention. Wang et al. \cite{wang2017residual} presented a residual attention network, which contains an attention module featuring an encoder-decoder architecture. Attention mechanism has also been utilized in semantic segmentation tasks to make more accurate dense predictions. For instance, Li et. al \cite{li2018pyramid} proposed a Pyramid Attention Network to exploit the impact of global contextual information in semantic segmentation.

These typical algorithms resemble in some aspects. Certain operations, such as global pooling, convolution and the combination of downsampling and upsampling, are utilized to enhance informative regions in the feature maps and suppress unrelated information, which makes the network learn more generalized visual structures and improves robustness to noisy inputs.

\subsection{Dilated Convolution}
Traditional deep convolutional networks often involve convolution with stride or pooling operations to improve receptive fields, and input images are downsampled in this process. However, these operations often lead to the loss of global information in dense predictions, such as semantic segmentation and object detection. Yu and Koltun \cite{yu2015multi} introduced dilated convolution to deep networks, which has proved useful in dense predictions. The basic idea of dilated convolution is to insert “holes” (zeros) in convolution kernels to obtain large receptive fields without downsampling. Dilated convolution avoids information loss during downsampling and has been widely used in semantic segmentation tasks \cite{wang2018smoothed} \cite{mehta2018espnet} \cite{park2018concentrated}. However, it has been observed that simply stacking dilated convolution in CNNs may cause grid effects and irrelevant long-ranged information \cite{yu2015multi} and lead to performance deterioration. Wang et. al \cite{wang2018understanding} proposed a hybrid dilated convolution (HDC) framework to avoid grid effects and improve segmentation performance on both large and tiny objects.

\subsection{AI-Based COVID-19 Segmentation Systems}
Artificial intelligence has been widely utilized in fighting against COVID-19. We mainly focus on AI-based semantic segmentation systems upon CT scans. Many works focus on learning robust and noise-insensitive representations from limited or noisy inputs. For example, Xie et. al \cite{xie2020relational} proposed a RTSU-Net for segmenting pulmonary lobes in CT scans. A non-local neural network module was introduced to learn both visual and geometric relationships among feature maps to produce self-attention. Wang et. al \cite{wang2020noise} presented a noise-robust framework for COVID-19 lesion segmentation. They utilized a noise-robust Dice loss and adaptive self-ensembling strategy to learn from noisy labels. Chen et. al \cite{chen2020residual} proposed a residual attention U-Net which introduced aggregated residual transformations and soft attention mechanism to learn robust feature representations. Also, researchers look into segmentation solutions that achieve both high speed and high accuracy. For example, Zhou et. al \cite{zhou2020rapid} developed a rapid, accurate and machine-agnostic segmentation and quantification method for automatic segmentation on COVID-19 lesions. The innovation of their work lies in the first CT scan simulator for COVID-19 and a novel network architecture which solves the large-scene-small-object problem. Qiu et. al \cite{qiu2020miniseg} developed a parameter-efficient framework to achieve fast segmentation of COVID-19 lung infection with relatively low computational cost.

\section{Methods}
\label{sec:methods}
In this section we will go through the details of the proposed \textit{D2A U-Net} architecture. In the first part, we will offer the overview of proposed network. We then provide details about dual attention mechanism and proposed attention modules. Finally we introduce our proposed decoder blocks.

\subsection{Overview of Network Architecture}
\label{sec:netarchitecture}
Basically, our proposed network is based on the U-Net \cite{ronneberger2015u} architecture, which is quite popular in medical image segmentation. Compared with original U-Net, dilated convolution and a novel combination of attention mechanism are integrated in our framework to obtain better feature representation. As COVID-19 pandemic broke out rapidly, available open access CT image data with gold-standard annotations is hard to acquire, and thus utilizing pretrained encoder in the segmentation model can offer a better parameter initialization and improve generalization ability. Therefore, in this work, we utilize a ResNeXt-50 (32$\times$4d) \cite{xie2017aggregated} pretrained on ImageNet as the encoder of our model. Furthermore, we integrate a dual attention mechanism in model decoder. A gated attention mechanism is inserted inside skip connections to utilize both high and low levels of feature representations and reduce semantic gap between encoder and decoder. Also, we introduce another fused attention mechanism in model decoder to refine feature maps after upsamling. We utilize a hybrid dilated convolution module \cite{wang2018understanding} as the basic block of model decoder to enlarge receptive field and produce better dense predictions. The network scheme is shown in Fig. \ref{fig:netimage}.

\subsection{Dual Attention Mechanism}
We introduce a \textit{dual attention mechanism} composed of a \textit{gate attention module} (GAM) and a \textit{decoder attention module} (DAM) to our network. GAM is utilized to refine features extracted by model encoder and reduce semantic gap by fusing high and low level feature maps. DAM is inserted in model decoder to refine feature representations after upsampling.

\subsubsection{Gate Attention Module}
\label{sec:gateattention}

\begin{figure*}[t] 
	\centering 
	\includegraphics[width=0.9\textwidth]{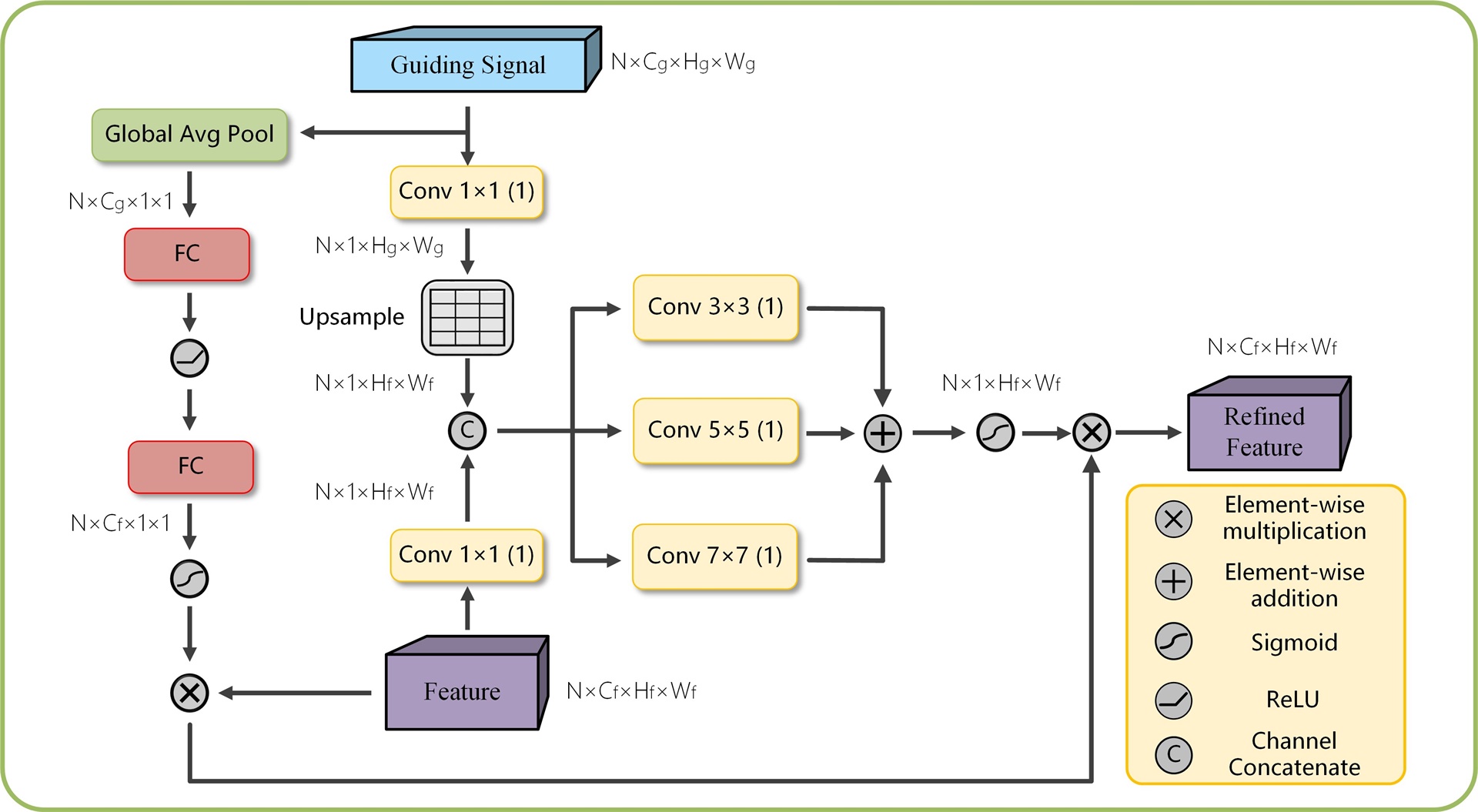} 
	\caption{The proposed \textit{gate attention module}, which takes guiding signal and features as input to generate fused attention. The number shown in the parentheses inside conv block means the number of outchannels. See \ref{sec:gateattention} for details.} 
	\label{fig:gateattention} 
\end{figure*}

Feature concatenation from encoder to decoder is the typical topological structure in U-Net, where the combination of high-resolution features in the encoder and upsampled features in the decoder enables better localization of segmentation targets \cite{ronneberger2015u}. However, not all visual representations in encoder feature maps contribute to precise segmentation. Also, semantic gap between encoder and decoder could limit model performance as well. Thus, we introduce a \textit{gate attention module} before concatenation to refine features coming from model encoder and reduce semantic gap.

Oktay et. al \cite{oktay2018attention} proposed an attention gate to refine encoder features with attention mechanism. But in their proposed attention gate only spatial attention mechanism is implemented to refine features. We believe introducing channel attention and spatial attention simultaneously will improve the efficiency of attention mechanism. Thus, inspired by the global attention upsample module proposed in pyramid attention network \cite{li2018pyramid} and CBAM \cite{woo2018cbam}, we provide a novel design of a \textit{gate attention module} to enable both channel attention and spatial attention. Detailed scheme of the proposed GAM is shown in Fig. \ref{fig:gateattention}. Two feature maps are fed into the attention module. The guiding signal refers to the feature map coming from model decoder (or the last convolution block in model encoder), and the feature refers to feature maps coming from model encoder to concatenate with upsampled feature maps. We use $\mathbf{G} \in \mathbb{R}^{C_g \times H_g \times W_g}$ to denote guiding signal and $\mathbf{F} \in \mathbb{R}^{C_f \times H_f \times W_f}$ to denote features.

In a U-Net shaped architecture, compared with $\mathbf{F}$, $\mathbf{G}$ contains more deep and high-resolution semantic information which is encoded in channel dimension. We utilize global average pooling and a multilayer perception (MLP) to create a channel attention map $Z_c(\mathbf{F}) \in \mathbb{R}^{C_f \times 1 \times 1}$. The output size of the MLP is smaller than the input size, thus we suppress irrelevant feature representations in channel dimension and implement channel-wise attention mechanism. In short, we compute channel attention as follows:

\begin{equation}
\begin{split}
Z_c(\mathbf{F})  &= \sigma (MLP(P_{avg}(\mathbf{G}))) \\
&= \sigma(W_{C_f}(ReLU(W_{C_g / r}(P_{avg}(\mathbf{G}))))) \\
\end{split}
\end{equation}

\noindent where $\sigma$ denotes sigmoid activation, $P_{avg}$ denotes global average pooling, $W_0 \in \mathbb{R}^{C_g / r \times C_g}$ and $W_1 \in \mathbb{R}^{C_f \times C_g / r} $, $r$ denotes reduce ratio and in our experiments it is set to 16.

Spatial attention is guided by both guiding signal and input feature itself. We use convolution operation with 1 filter to squeeze channel dimension of $\mathbf{G}$ and $\mathbf{F}$. Then reduced feature map from $\mathbf{G}$ is upsampled to match the size of $\mathbf{F}$. A combination of convolution operation with different kernel size is utilized to produce spatial attention $Z_s(\mathbf{F}) \in \mathbb{R}^{1 \times H_f \times W_f}$. In short, we compute spatial attention as:

\begin{equation}
\begin{split}
& Z_s(\mathbf{F})  = \sigma (f_{3 \times 3}([\mathbf{F_r, \mathbf{G_r}}])+f_{5 \times 5}([\mathbf{F_r, \mathbf{G_r}}])+f_{7 \times 7}([\mathbf{F_r, \mathbf{G_r}}])) \\
& where \quad \mathbf{F_r} = f^r_{1\times1}(\mathbf{F}), \ \mathbf{G_r} = upsample(f^r_{1\times1}(\mathbf{G}))
\end{split}
\end{equation}

\noindent where $\sigma$ denotes sigmoid activation, $f_{3\times3}$, $f_{5\times5}$ and $f_{7\times7}$ denote convolution operation with corresponding kernel size. $f^r_{1\times1}$ is used to squeeze channel dimension.

Then we use element-wise multiplication to combine spatial and channel attention to produce fused attention $Z(\mathbf{F})$:

\begin{equation}
Z(\mathbf{F})=\mathbf{F}*Z_s(\mathbf{F})*Z_c(\mathbf{F})
\end{equation}

\subsubsection{Decoder Attention Module}
\label{sec:dam}

\begin{figure*}[t] 
	\centering 
	\includegraphics[width=0.9\textwidth]{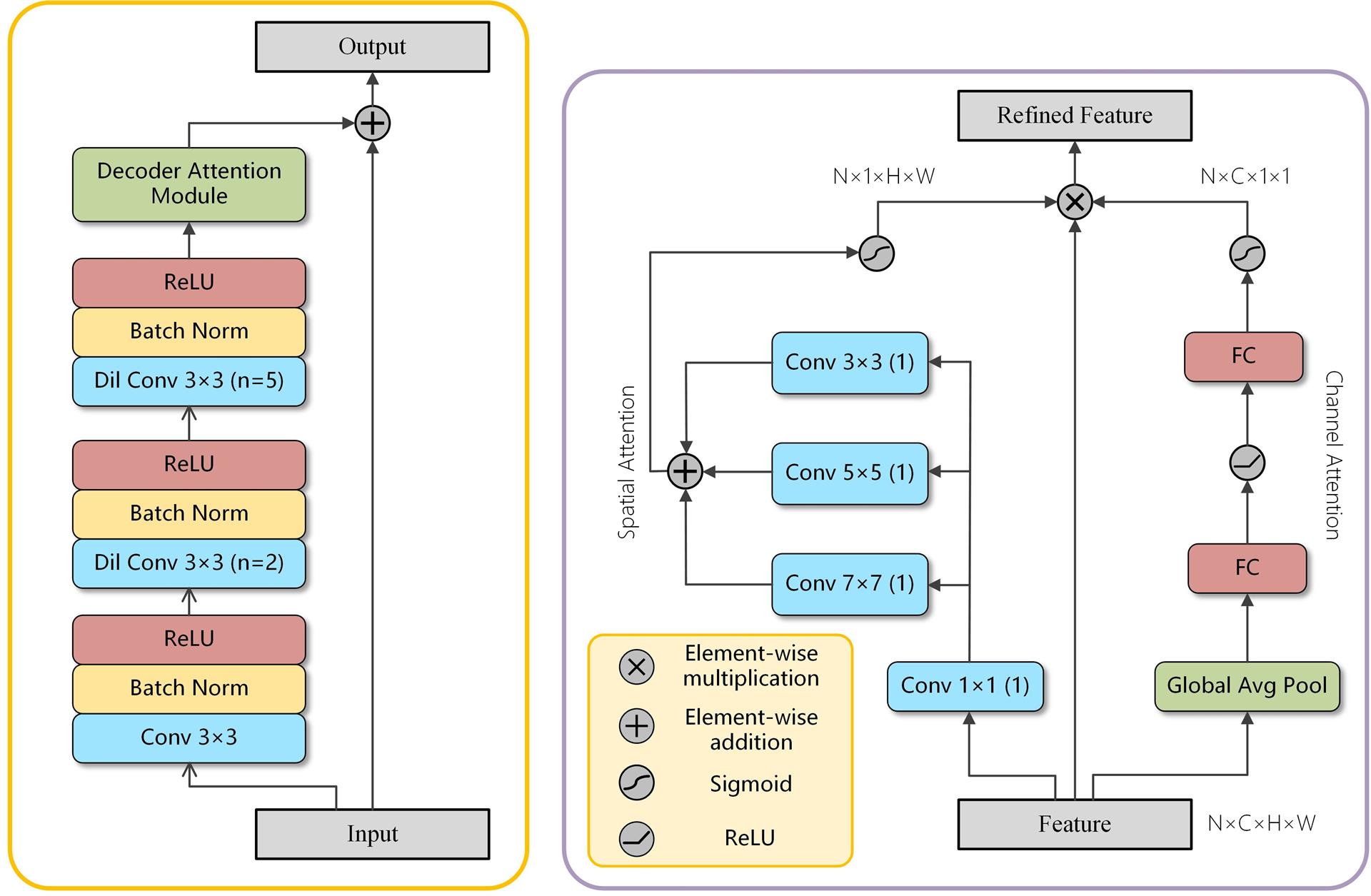} 
	\caption{The proposed \textit{residual attention block} (left) and \textit{decoder attention module} (right). RAB integrates a hybrid dilated convolution module and a DAM; $n$ in the parentheses refers to dilation rate. DAM is utilized to refine post-upsample features; the number shown in the parentheses inside conv block means the number of outchannels. See \ref{sec:rablock} for details about RAB and \ref{sec:dam} for details about DAM.} 
	\label{fig:rablock} 
\end{figure*}

In semantic segmentation, high-resolution visual representations in the encoder need to be upsampled to make dense predictions. Transposed convolution and interpolation are both popular solutions to image upsampling, but both have their drawbacks. Compared with interpolation, transposed convolution is trainable and offers more nonlinearity to deep networks, which improves model fitting capacity. But grid effect is hard to avoid if hyperparameters are not configured properly, while such drawback can be more troublesome when stacking more than one transposed convolution layer. Thus we propose a combination of bilinear interpolation and following convolution to upsample feature maps. However, as interpolation is not trainable, it is inevitable to introduce irrelevant information or noise to upsampling. We introduce a \textit{decoder attention module} to solve this issue. A fused attention mechanism is utilized to refine post-upsampling feature maps in both channel and spatial dimensions. The scheme is shown in Fig. \ref{fig:rablock}. Compared with GAM, DAM is more simplified and only takes one input, but the implementation of both channel and spatial attention is quite similar. We use $Z_c(\mathbf{F}) \in \mathbb{R}^{C \times 1 \times 1}$ to denote channel attention, $\ Z_s(\mathbf{F}) \in \mathbb{R}^{1 \times H \times W}$ to denote spatial attention and $Z(\mathbf{F})$ to denote fused attention. In short, DAM is computed as follows:

\begin{equation}
\begin{split}
Z_c(\mathbf{F})  &= \sigma (MLP(P_{avg}(\mathbf{F}))) \\
&= \sigma(W_1(ReLU(W_0(P_{avg}(\mathbf{F})))))
\end{split}
\end{equation}

\noindent where $\sigma$ denotes sigmoid activation, $P_{avg}$ denotes global average pooling, $W_0 \in \mathbb{R}^{C / r \times C}$ and $W_1 \in \mathbb{R}^{C \times C / r} $, $r$ denotes reduce ratio and in our experiments it is set to 16.

\begin{equation}
Z_s(\mathbf{F})  = \sigma (f_{3\times3}(f^r_{1\times1}(\mathbf{F}))+f_{5\times5}(f^r_{1\times1}(\mathbf{F}))+f_{7\times7}(f^r_{1\times1}(\mathbf{F}))
\end{equation}

\noindent where $\sigma$ denotes sigmoid activation, $f_{3\times3}$, $f_{5\times5}$ and $f_{7\times7}$ denote convolution operation with corresponding kernel size. $f^r_{1\times1}$ is used to squeeze channel dimension.

\begin{equation}
Z(\mathbf{F})=\mathbf{F}*Z_s(\mathbf{F})*Z_c(\mathbf{F})
\end{equation}

\renewcommand{\arraystretch}{1.6}
\begin{table*}[h]
	\caption{Dataset Description}
	\centering
	\label{table:dataset}
	\begin{tabular}{|c|c|c|c|}
		\hline
		Num & Dataset & Description & Split \\ 
		\hline
		1 & \makecell*[c]{COVID-19 CT segmentation dataset\cite{covid19ctseg}} & 110 slices with 100 containing annotations. & Test Set \\ 
		\hline
		2 & \makecell*[c]{Segmentation dataset nr. 2\cite{covid19ctseg}} & \makecell*[c]{9 CT volumes (373 out of the total of 829 slices \\have been evaluated by a radiologist as positive \\and segmented.)} & Training Set \\ 
		\hline
		3 & \makecell*[c]{COVID-19 CT Lung and Infection \\ Segmentation Dataset\cite{ma_jun_2020_3757476}} & \makecell*[c]{20 CT volume (Left lung, right lung, and infections \\are labeled by two radiologists and verified by an \\experienced radiologist, and 1,844 out of the total of \\3520 slices contains infection regions.)} & Training Set \\ 
		\hline
	\end{tabular}
\end{table*}

\subsection{Residual Attention Block}
\label{sec:rablock}
Standard convolution hardly reaches a large receptive field due to kernel size. Such drawback in traditional design of U-Net based network decoder can limit the performance in segmentation. Inspired by the design of hybrid dilated convolution \cite{wang2018understanding}, we proposed a \textit{residual attention block} (RAB) as the basic module in model decoder. Unlike similar works using dilated convolution in the encoder, we explore to use it in the decoder to capture multiscale patterns of upsampled feature maps. Hybrid dilated convolution is utilized in our RAB to acquire large receptive fields and avoid grid effects. The stem of RAB is a stack of dilated convolution with kernel size 3 and dilation rate [1, 2, 5], followed by a \textit{decoder attention module}. The scheme is shown in Fig. \ref{fig:rablock}.

We assume initial receptive field as 1 $\times$ 1. The equivalent kernel size of dilated convolution is computed as follows:

\begin{equation}
K = k + (k - 1)(n - 1)
\end{equation}

\noindent where $K$ denotes equivalent kernel size, $k$ denotes actual kernel size and $n$ denotes dilation rate.

Thus, the equivalent kernel size of dilated convolution with kernel size 3 and dilation rate [1, 2, 5] is 3, 5, 11, respectively. According to the definition of receptive field, such design of stacked dilated convolution reaches a receptive field of 17 $\times$ 17, which enables the capture of global information. Also, dilated convolution with different dilation rate can capture multiscale information in feature maps, which can contribute to the accurate segmentation on both large and small objects.

As we use a ResNeXt-50 (32 $\times$ 4d) as model encoder, we utilize residual connection in decoder as well to avoid gradient vanishing. Dilated convolution is followed by a DAM to refine upsampled features and produce fused attention maps. In short, the output of our RAB is computed as follows:

\begin{equation}
\mathbf{Y} = \mathbf{X} + DAM(HDC(\mathbf{X}))
\end{equation}

\noindent where $DAM$ denotes \textit{decoder attention module} and $HDC$ denotes hybrid dilated convolution.

\section{Experiments}
\label{sec:experiments}

\subsection{CT Segmentation Dataset}
CT slices used in our experiments consist of 3 datasets\cite{covid19ctseg}\cite{ma_jun_2020_3757476}. Details about dataset used are shown in Table \ref{table:dataset}. Dataset 1 contains 100 axial CT slices from more than 40 patients, which have been rescaled to 512 $\times$ 512 pixels and grayscaled. All slices are segmented by a radiologist using three labels: ground-glass opacity, consolidation and pleural effusion. Dataset 2 contains 9 axial CT volumes, where 373 out of the total of 829 slices have been evaluated by a radiologist as positive and segmented using 2 labels including ground-glass opacity and consolidation. Dataset 3 contains 20 CT axial volumes, which have been segmented by two radiologists and verified by an experienced radiologist.

Dataset 2 and Dataset 3 contain 29 CT volumes in total, but not all slices contain infection regions. We choose to discard all slices containing no COVID-19 infection and use slices with annotations only. As annotations in Dataset 3 do not distinguish ground-glass opacity and consolidation, we take both ground-glass opacity and consolidation in Dataset 2 as COVID-19 lesions and do not distinguish them as well, thus creating a binary segmentation dataset. An intensity normalization has been applied on both datasets and all slices have been rescaled to 512 $\times$ 512 pixels to match Dataset 1. We take all ground-glass, consolidation and pleural effusion in Dataset 1 as COVID-19 lesions, just the same as what we have done to Dataset 2.

We did not choose to combine processed Dataset 1 to 3 together and then split them randomly, because in this way slices of one certain subject can exist in both training and test datasets, which could be regarded as a data leakage and cause a virtual-high model performance. Instead, we finally obtain 1645 processed slices from processed Dataset 2 and Dataset 3 in total and use these slices as our final training dataset, and then we use 100 axial slices from Dataset 1 as our final test dataset. Such data split can best evaluate model generalization capacity.

\subsection{Implementation Details}

\subsubsection{Model Hyperparameters and Settings}

Model encoder is a ResNeXt-50 (32 $\times$ 4d) pretrained on ImageNet-1K. We remove the global average pooling and full connection layers from original network. Number of output channels is 64, 256, 512, 1024, 2048, respectively, just the same as original paper of ResNeXt. Convolution operations in model decoder are padded, without stride, if not specified. Bilinear interpolation is utilized to upsample feature maps, and scale factor is set to 2. Dice loss is widely utilized in semantic segmentation, but the differential of Dice loss is sometimes numerically unstable and may lead to oscillation in training process. The combination of Dice loss and cross-entropy could avoid this issue. Thus we combine Dice loss $\mathcal{L}_d$ and binary cross-entropy loss $\mathcal{L}_c$ as our final loss function:

\begin{equation}
\mathcal{L}_{seg} = \mathcal{L}_d + \alpha \mathcal{L}_c
\end{equation}

\noindent where $\alpha = 1$ in our experiments.

\subsubsection{Training Details}

Our model is implemented using PyTorch on an Ubuntu 16.04 server. We use a NVIDIA RTX 2080 Ti GPU to accelerate our training process. Data augmentation is utilized in our training process to reduce overfitting and improve generalization capacity. First all input images are rescaled to 560 $\times$ 560, followed by random flip, random rotation, random gamma and log transform. Finally images are randomly cropped to 448 $\times$ 448 and fed into network. The model is optimized by an Adam optimizer with $\beta_1 = 0.9$, $\beta_2 = 0.999$, $\epsilon = 1e-8$. $L_2$ regularization is utilized to reduce overfitting as well. We set model weight decay to 1e-4. Initial learning rate is set to 1e-4 and reduced when faced with plateau, with reduce factor being 0.1 and patience being 10. The batch size is set to 6 and we perform evaluation on test set after 30 epochs. The training process takes approximately 140 minutes.

\subsection{Evaluation Metrics}
We use Dice similarity coefficient and pixel error as the main metrics to evaluate segmentation performance of our \textit{D2A U-Net}. Dice is a statistic used to gauge the similarity of two samples, and has been widely used to evaluate performance in semantic segmentation. Pixel error measures the number of pixels predicted falsely in the image, which shows the global segmentation accuracy of the proposed models. Both metrics measure segmentation performance in a global way. In addition, we calculate recall score of infection regions as recall score measures model's sensitivity to lung infection, which is rather significant in terms of COVID-19 infection. We use $G$ to denote ground truth, $P$ to denote dense predications, $TP$ to denote true positive, $FP$ to denote false positive, $TN$ to denote true negative and $FN$ to denote false negative. These metrics are calculated as follows:

\begin{equation}
\begin{split}
Dice &= \frac{2 |G \bigcap P|}{|G| + |P|} \\
&= \frac{2TP}{2TP + FP + FN}
\end{split}
\end{equation}

\begin{equation}
Pixel \ Error = \frac{FP + FN}{TP + TN + FP + FN}
\end{equation}

\begin{equation}
Recall = \frac{TP}{TP + FN}
\end{equation}

\subsection{Comparison with Cutting-Edge Methods}

We compared the performance of proposed network with U-Net \cite{ronneberger2015u}, Attention U-Net \cite{oktay2018attention} and U-Net$++$ \cite{zhou2018unet++}. The VGG-style backbone refers to the encoder design proposed in original U-Net paper \cite{ronneberger2015u}.

Also, we compared our model with 2 cutting-edge models widely used in natural image segmentation, including FCN8s \cite{long2015fully} and DeepLab v3 (output stride = 8) \cite{chen2017rethinking}, with both models containing a pretrained backbone as well.

Apart from model performance comparison, model parameters and computational costs (FLOPs) are also compared in our experiments.

As our model differs with other U-Net family models in terms of model encoder, to best evaluate our design of model decoder and attention mechanism, we also build a simplified \textit{D2A U-Net} with a VGG-style backbone as well. We believe the simplified version offers more fair comparison between proposed network and other U-Net based models, while standard \textit{D2A U-Net} with backbone ResNeXt-50 (32 $\times$ 4d) provides best segmentation results.

To best evaluate model performance, all the metrics reported in Table \ref{table:quantitative} are averaged in 5 reduplicate experiments to report a fair and reliable result.

\subsection{Segmentation Results}

\subsubsection{Quantitative Analysis}
Detailed comparison among different models in our experiments is shown in Table \ref{table:quantitative}. As has been shown, without pretrained backbone, our proposed network outperforms U-Net, Attention U-Net and U-Net$++$ in terms of Dice, pixel error and recall. As these models are identical in model encoder, it is clear that the proposed dual attention mechanism and RAB contribute to infection segmentation a lot. The utilization of attention mechanism aids the model to detect infected tissues more accurately, which reduces the number of false positives and improves recall score. Also, RAB in model decoder captures both large and tiny visual structures, which is helpful to segment infection lesions with different size. Also, it should be noted that proposed \textit{D2A U-Net} with VGG-style backbone outperforms U-Net$++$ with comparably lower model parameters and computational costs, which could prove the balance of efficiency and performance in our models. 

Utilizing pretrained backbone could also improve model performance. As can be seen, our \textit{D2A U-Net} with pretrained ResNeXt-50 (32 $\times$ 4d) backbone outperforms other networks including ones with similar pretrianed backbones in terms of Dice, pixel error and recall by a large margin and yields best results on our dataset. Also, our \textit{D2A U-Net} with pretrained ResNeXt-50 (32 $\times$ 4d) backbone takes fewer computational resources than FCN8s and DeepLab v3 (output stride = 8). As can be seen from Table \ref{table:quantitative}, pretrained encoder could offer a better initialization of model parameters and reduce overfitting, especially when data amount is insufficient. Overall, the proposed architecture performs better than existing cutting-edge models. 

\begin{table*}[h]
	\caption{Quantitative analysis of infection regions on our dataset. Backbone \textit{VGG-style} refers to the encoder proposed in \cite{ronneberger2015u}, and backbone \textit{ResNet-101} and \textit{ResNeXt-50 (32 $\times$ 4d)} are pretrained on ImageNet-1K.} 
	\centering
	\label{table:quantitative}
	\setlength{\tabcolsep}{2mm}{
		\begin{tabular}{@{}l|lll|lll@{}}
			\toprule
			Model              & Backbone & Param. & FLOPs & Dice     & Pixel Error           & Recall          \\ \midrule
			U-Net              & VGG-style   & 7.85 M & 43.13 G   & 0.6384          & 0.0332        & 0.5512          \\
			Attention U-Net    & VGG-style   & 8.12 M & 43.78 G   & 0.6646          & 0.0390        & 0.6470           \\
			U-Net++            & VGG-style  & 9.16 M & 106.81 G    & 0.6830         & 0.0332         & 0.6417          \\
			DeepLab v3 (os=8)         & ResNet-101  & 58.63 M & 185.00 G   & 0.7095         & 0.0323         & 0.6780 \\
			FCN8s                & ResNet-101  & 51.94 M & 165.67 G   & 0.6825          & 0.0315        & 0.6348 \\ \bottomrule
			\textit{D2A U-Net} & VGG-style  & 8.95 M & 53.19 G    & 0.7047        & 0.0323          & 0.6626 \\ 
			\textit{D2A U-Net} & ResNeXt-50 & 90.05 M & 149.97 G & \textbf{0.7298} & \textbf{0.0311}   & \textbf{0.7071} \\ \bottomrule
	\end{tabular}}
\end{table*}

\subsubsection{Qualitative Analysis}

\begin{figure*}[h] 
	\centering 
	\includegraphics[width=1\textwidth]{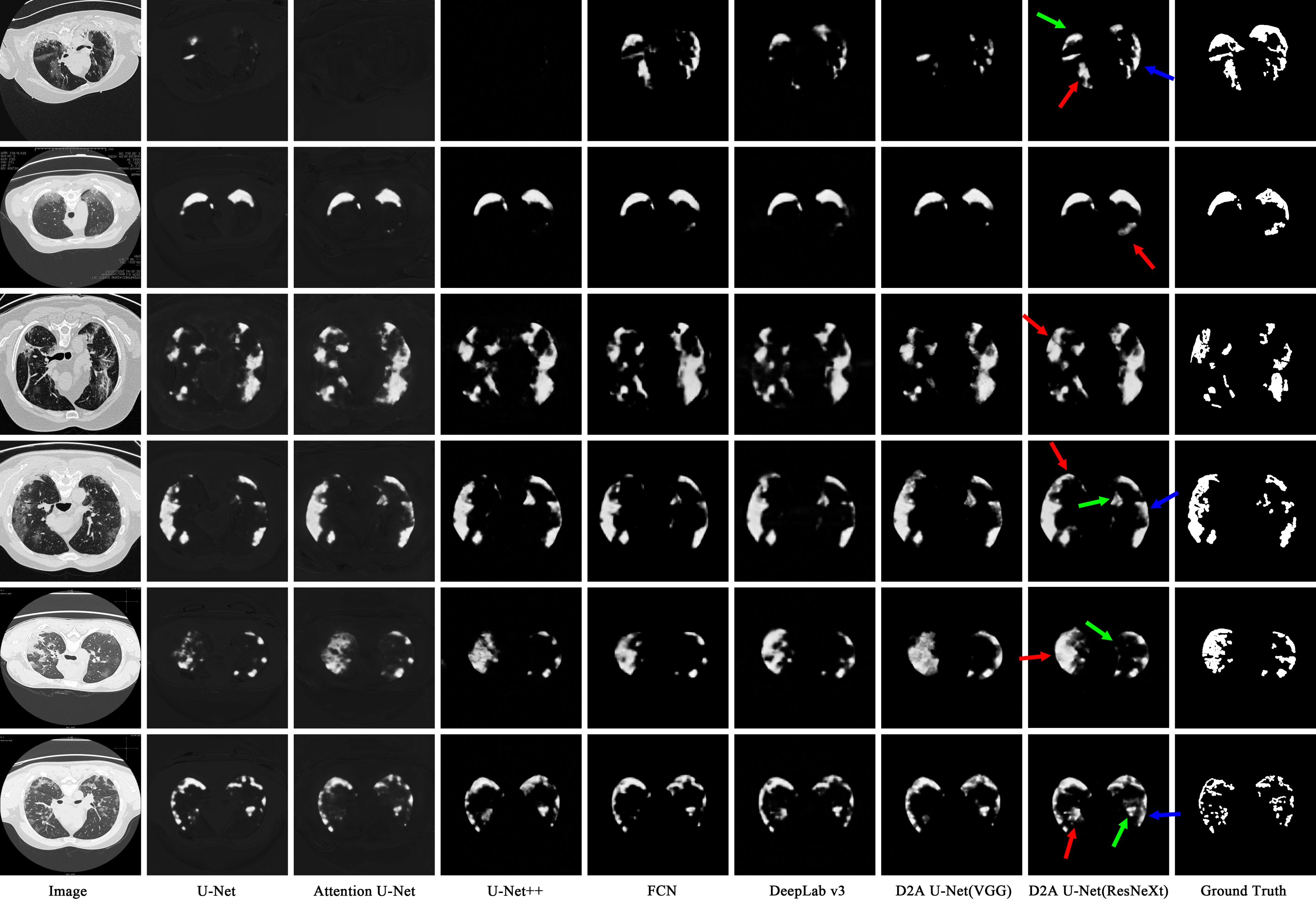} 
	\caption{Visual comparison of COVID-19 lesions segmentation results.} 
	\label{fig:visualization} 
\end{figure*}

We visualized segmentation results, as is shown in Fig. \ref{fig:visualization}. It can be seen from the visualization that our proposed model outperforms other models obviously. U-Net and Attention U-Net are the least sensitive to COVID-19 lesions, and the background pixels have much stronger activation compared with other models. U-Net$++$ produces more accurate segmentation results, but still not promising as some tiny lesions or lesions with blurred edge are segmented poorly. \textit{D2A U-Net} with VGG-style backbone produces most accurate segmentation masks compared with other U-Net based models mentioned above, and when backbone is switched to ResNeXt-50 (32 $\times$ 4d), \textit{D2A U-Net} produces the best segmentation results, which is comparably more sensitive to blurred or tiny lesions than other models.

\subsubsection{Ablation Study}

Several ablation experiments were conducted to evaluate the performance of components presented in our model, as is shown in Table. \ref{table:ablationanalysis}. 

\paragraph{Effectiveness of Proposed GAM}
\label{para:GAM}

To evaluate the validity of proposed GAM in our experiments, we designed two baselines shown in Table. \ref{table:ablationanalysis}, including No.1 (U-Net only) and No.2 (U-Net + GAM). Experimental results have shown that introducing GAM to U-Net model can boost the performance, which leads to a better Dice score and recall.

\paragraph{Effectiveness of Proposed RAB}
\label{para:RAB}
We conducted similar experiments (No.1 and No.3) to explore the effectiveness of proposed RAB, which includes a hybrid dilated convolution block and a decoder attention module. Experimental results indicate that introducing RAB to our model yields better results as well, but the performance boost is comparably limited compared with GAM.

\paragraph{Effectiveness of Combining GAM, RAB and PB}
As can be seen from Table. \ref{table:ablationanalysis}, in experiment No.4, introducing GAM and RAB together (proposed \textit{D2A U-Net}) yields best results in our experiments, and the performance boost exceeds the simple addition of each module's performance boost. Such experimental results indicate that introducing GAM and RAB together promotes the performance mutually. Also, in No.5, pretrained backbone as better parameter initialization could further improve model performance.

\begin{table*}[h]
	\caption{Ablation analysis of proposed \textit{D2A U-Net}, where GAM denotes \textit{gate attention module}, RAB denotes \textit{residual attention block} and PB denotes pretrained backbone.}
	\centering
	\label{table:ablationanalysis}
	\setlength{\tabcolsep}{2mm}{
		\begin{tabular}{@{}lllll@{}}
			\toprule
			Method                   & Dice & Pixel Error      & Recall \\ \midrule
			(No.1) U-Net             & 0.6384      & 0.0332      & 0.5512 \\
			(No.2) U-Net + GAM       & 0.6771      & 0.0343      & 0.6445 \\
			(No.3) U-Net + RAB       & 0.6579      & 0.0354      & 0.6154 \\
			(No.4) U-Net + RAB + GAM & 0.7047      & 0.0323      & 0.6626 \\ \bottomrule
			(No.5) U-Net + RAB + GAM + PB & \textbf{0.7298} & \textbf{0.0311} & \textbf{0.7071} \\ \bottomrule
	\end{tabular}}
\end{table*}

\section{Conclusion}
\label{sec:conclusion}
In this paper we proposed a novel segmentation network, \textit{D2A U-Net}, for COVID-19 CT segmentation. Inspired by global attention upsample and CBAM, we propose a novel gated attention mechanism, called \textit{gate attention module}, to produce a fused attention map on features extracted by encoder. We introduce a \textit{decoder attention module} as well, which helps refine upsampled feature maps. Also, inspired by hybrid dilated convolution, we present a \textit{residual attention block} containing a hybrid dilated convolution and a \textit{decoder attention module}; we use it as the basic block in model decoder. Attention mechanism is utilized to increase model sensitivity to positive pixels and improve recall score. And we use residual attention block as decoder basic block to refine upsampled feature maps and increase receptive field simultaneously. Experimental results indicate that our network design is capable of segment COVID-19 lesions from CT slices automatically, and achieves best results among popular cutting-edge models evaluated in our experiments. But our work is still limited to some degree, as only binary segmentation is performed in our experiments, which can limit model’s potential use in both diagnosis and health care. We expect to gather more CT scans and perform multi-class segmentation in the future. Also, despite the significantly better performance of our \textit{D2A U-Net} with ResNeXt-50 (32 $\times$ 4d) backbone, the model has much more model parameters than other architectures with similar backbones (FCN8s and DeepLab v3). We believe that as ResNet family models have a large number of channels (eg. 1024 and 2048 in the last two layers), the parameters of model decoder becomes extremely large. Such problem might be addressed by introducing so-called Bottleneck in ResNets to the decoder of \textit{D2A U-Net} to reduce channels and thus model parameters.

\section*{Acknowledgements}
This work was partially supported by the Fundamental Research Funds for Central Universities, the National Natural Science Foundation of China (No. 61601019, 61871022), the Beijing Natural Science Foundation (7202102), and the 111 Project (No. B13003).


\bibliographystyle{cas-model2-names}

\bibliography{references}

\end{document}